\documentclass[%
twocolumn,
 amsmath,amssymb,
 prl,
]{revtex4-1}
\usepackage{lineno}
\usepackage{xcolor}

\usepackage{graphicx}
\usepackage{dcolumn}
\usepackage{bm}
\begin{document}

\preprint{}

\title{Radiation reaction cooling as a source of anisotropic momentum distributions with inverted populations}

\author{P. J. Bilbao}%
 \email{pablojbilbao@tecnico.ulisboa.pt}
\author{L. O. Silva}%
 \email{luis.silva@tecnico.ulisboa.pt}
\affiliation{%
 GoLP/Instituto de Plasmas e Fus\~{a}o Nuclear, Instituto Superior T\'{e}cnico, Universidade de Lisboa, 1049-001 Lisbon, Portugal
}%

\date{Submitted: 26th October 2022, Accepted: 28th March 2023}

\begin{abstract}
Under the presence of strong electromagnetic fields and radiation reaction, plasmas develop anisotropic momentum distributions, characterized by a population inversion. This is a general property of collisionless plasmas when the radiation reaction force is taken into account. We study the case of a plasma in a strong magnetic field and demonstrate the development of ring momentum distributions. The timescales for ring formation are derived for this configuration. The analytical results for the ring properties and the timescales for ring formation are confirmed with particle-in-cell simulations. The resulting momentum distributions are kinetically unstable and are known to lead to coherent radiation emission in astrophysical plasmas and laboratory setups. 
\end{abstract}

\maketitle
In the presence of strong fields, relativistic charged particles can radiate photons with energy comparable to the rest mass of the electron $m_e c^2$ or even comparable to the kinetic energy of the particle $(\gamma-1) m_e c^2$, where $\gamma$ is the Lorentz factor of the charged particle, $m_e$ is the electron mass. In these scenarios, radiation reaction \emph{i.e.} the momentum recoil due to the radiation emission, must be taken into account and it modifies the dynamics of relativistic charged particles \textcolor{black}{\cite{gonoskov2022charged}}. The conditions for radiation reaction to be important are present around compact objects \cite{kaspi2017magnetars, cerutti2017electrodynamics, xue2019magnetar}, in experiments with intense lasers \cite{di2009strong, di2012extremely, thomas2012strong, vranic2014all, vranic2016quantum_m}, magnetic field amplification laboratory scenarios \cite{nakamura2018record, jiang2021magnetic}, and fusion plasmas \cite{hirvijoki2015radiation, decker2016numerical}. The interplay between radiation reaction and global plasma dynamics has only very recently started to be addressed \textcolor{black}{\cite{grismayer2016laser,liseykina2016inverse, gong2019radiation, qu2021collective}}.

The radiation reaction force or radiation friction force (for a review see \textcolor{black}{\cite{gonoskov2022charged}}) does not conserve the momentum-space volume, unlike conservative forces like the Lorentz force. The impact of radiation reaction on the collective dynamics of plasmas was hinted in recent works on runaway electrons in fusion plasmas that have shown that radiation reaction and collisional effect induce ``bumps'' along the runaway electron tail \cite{hirvijoki2015radiation,   decker2016numerical}. In this Letter, we show that this behavior is more general and that anisotropic momentum distributions can be produced in other regions of the momentum distribution due to the properties of the radiation reaction force.

The radiation power due to synchrotron radiation is $\propto \gamma^4 (\mathbf{p} \times \mathbf{a})^2$, where $\mathbf{p}$ is the momentum of the particle, and  $\mathbf{a}$ is the acceleration \cite{lightman1982relativistic, rybicki1991radiative}; this already shows that different regions of momentum space cool at different rates due to radiation reaction, \emph{i.e.} differential cooling. 
In this work, we show how this differential cooling in momentum space will result in anisotropic regions in phase-space, with bunching in momentum space. This effect is a general feature of radiation reaction cooling and its importance depends on the specific details of the field configuration. In this work, we thoroughly analyze the simplest scenario (a plasma in a constant strong magnetic field), which captures the key features of this process.
This effect is analytically demonstrated by including the radiation reaction force into the Vlasov equation \cite{kuz1978bogolyubov,hazeltine2004radiation, tamburini2011radiation, hirvijoki2015radiation, decker2016numerical}; analytical results for the evolution of the distribution function, and the relevant timescales for the resulting momentum distribution are derived for this configuration.
We show that the resulting momentum distributions always develop an inverted Landau population, \emph{i.e.} a region of the momentum distribution $f (\mathbf{p},t)$ that fulfils $\partial f/ \partial p_\perp > 0$, where $p_\perp$ refers to the momentum direction perpendicular to the magnetic field. Particle-in-cell (PIC) simulations confirm the theoretical results for a broad range of initial conditions and in both the classical $\chi \ll 1$ and in the quantum regimes $\chi > 1$ ($\chi$ is the Lorentz- and Gauge-invariant parameter $\chi=e\sqrt{-(F_{\mu\nu} p^\nu)^2}/m_e^3$ \cite{ritus1985quantum, di2012extremely}; $e$ is the electron charge, $F_{\mu\nu}$ the electromagnetic tensor and $p^\nu$ the 4-momentum of the particle. For a constant background magnetic field, $\chi$ reduces to  $\chi=p_\perp \left|\mathbf{B_0}\right| /(m_e B_{Sc})$, where $B_{Sc} = m_e^2 c^2 /(e \hbar) \simeq 4.41\times10^9$ T is the Schwinger critical field.).

The resulting momentum distributions with inverted Landau populations are known to be kinetically unstable and responsible for providing the free energy for kinetic plasma instabilities and coherent radiation mechanisms such as the electron cyclotron maser instability \cite{le1984direct,bingham2000generation, cairns2005cyclotron, cairns2008cyclotron, melrose2016cyclotron}. Thus, radiation reaction naturally leads to the conditions required for the seeding of instabilities and coherent radiation driven by inverted Landau populations and we explore how these results are relevant for astrophysical and laboratory plasmas \cite{melrose1995models, lyutikov1999nature, davoine2018ion,melrose2021pulsar, lyutikov2021coherent}.

We will first consider the classical regime of radiation reaction \cite{dirac1938classical,landau1975classical, hartemann1995classical, bell2008possibility, sokolov2009dynamics}. For $\gamma \gg 1$, the dominant contribution is the leading order in $\gamma$ and we can approximate the standard Landau-Lifshitz formula for radiation reaction as \cite{landau1975classical,tamburini2010radiation}
\begin{equation}
     \mathbf{F}_{rad} = -\frac{2}{3}\frac{e \alpha \gamma\mathbf{p}}{B_{Sc} m_ec^2}  \left[ \left( \mathbf{E} + \frac{\mathbf{p}}{\gamma m_e} \times \mathbf{B}\right)^2 - \left( \frac{\mathbf{p}}{\gamma m_e c } \cdot \mathbf{E} \right)^2\right], \label{eq:landau}
\end{equation}
where $\alpha$ is the fine-structure constant, and $\textbf{E}$ and $\textbf{B}$ are arbitrary electric and magnetic fields, respectively. Equation (\ref{eq:landau}) shows there is a non-linear dependence of the momentum radiation rate on the momentum of the particle and its acceleration, \emph{i.e.} the differential cooling. We note the implications of the differential cooling have not been examined in the context of deformations and bunching in momentum space \cite{commentemmittance,lee1982radiation, barletta1987linear}. As we are examining the case of plasmas in a constant magnetic field, the terms proportional to $\textbf{E}$ in Eq. (\ref{eq:landau}) are discarded, and we can take advantage of the cylindrical symmetry imposed by the  constant magnetic field. The momentum $\textbf{p}$ is decomposed into the parallel $p_\parallel$ and the perpendicular $p_\perp$ components with respect to $\textbf{B}$, such that $\left(\textbf{p} \times \textbf{B} \right)^2 = p_\perp^2 B^2$. We normalize the magnetic field $B$ as $B_0 = B/B_{Sc}$, $t$ to the inverse of the cyclotron frequency $\omega_{ce}^{-1} = m_e /(e B_0)$, the momentum to $m_ec$, such that $\gamma^2 = 1+\mathbf{p}^2$. Thus, $\chi$ and $\mathbf{F}_{rad}$ are given by $\chi=p_\perp B_0$, and 
$  \mathbf{F}_{rad}  = - \frac{2}{3} \alpha B_0 p_\perp ^2 \mathbf{p}/\gamma  $, respectively. 

Generalized kinetic equations for non-conservative forces, in particular for radiation reaction, have been known since the 1960s \cite{hakim1968relativistic, hakim1971collective, kuz1978bogolyubov}. We resort to the non-manifestly covariant form of the Vlasov equation including radiation reaction \cite{kuz1978bogolyubov, hazeltine2004radiation, hazeltine2004radiation, tamburini2011radiation, hirvijoki2015radiation, decker2016numerical}
\begin{equation}
    \frac{\partial f}{\partial t}+ \frac{\mathbf{p}}{m} \cdot \mathbf{\nabla}_\mathbf{r} f+ \mathbf{\nabla}_\mathbf{p} \cdot \left( \left( \mathbf{F}_{rad}+ \mathbf{F}_{L} \right) f \right) = 0, \label{eq:pre_Vlasov}
\end{equation}
where $f (\mathbf{p},\mathbf{r},t)$ is the distribution function, and $\mathbf{F}_{L}$ is Lorentz force. $\mathbf{\nabla}_\mathbf{r} f$ and $\mathbf{\nabla}_\mathbf{p} \cdot \left(\textbf{p} f\right)$ are the spatial gradient of $f$ and momentum divergence of $\textbf{p} f$, respectively. The inclusion of the radiation reaction force as the operator $\mathbf{\nabla}_p \cdot \left( \mathbf{F}_{rad} f \right)$ guarantees the conservation of the number of particles \cite{stahl2015effective}. Since $\mathbf{F}_{L}$ is conservative, but $\mathbf{F}_{rad}$ is dissipative, then $\mathbf{\nabla}_p \cdot \mathbf{F}_{rad} \ne\mathbf{\nabla}_p \cdot \mathbf{F}_{L} = 0$. For a spatially homogeneous plasma, we can neglect the term proportional to $\mathbf{\nabla}_\mathbf{r} f$. Moreover, as we are assuming cylindrical symmetry, the effect of the Lorentz force due to a strong magnetic field on the distribution is $\mathbf{\nabla}_p \cdot \left(  \mathbf{F}_{L} f \right) = 0$, even when $\mathbf{F}_{L} \ne 0$. Thus, Eq. (\ref{eq:pre_Vlasov}) is simplified to
\begin{equation}
    \frac{\partial f}{\partial t}+ \mathbf{F}_{rad}\cdot \mathbf{\nabla}_p f + f \mathbf{\nabla}_p \cdot\mathbf{F}_{rad} = 0. \label{eq:modVlasov_reduced}
\end{equation}
In cylindrical coordinates, the operators are $\mathbf{\nabla}_p f = \partial f/\partial p_\parallel \hat{\mathbf{p_\parallel}}+ \partial f/\partial p_\perp \hat{\mathbf{p_\perp}}$ and $\mathbf{\nabla}_\mathbf{p} \cdot\mathbf{F} = \frac{1}{p_\perp}\partial \left( p_\perp F_\perp \right)/\partial p_\perp +\partial F_\parallel/\partial p_\parallel$, yielding
\begin{equation}
    \frac{3}{2 \alpha B_0 }\frac{\partial f}{\partial t}  = \eta \frac{p_\perp^2}{\gamma}f + \frac{p_\perp^3}{\gamma } \frac{\partial f}{\partial p_\perp} + \frac{p_\perp^2 p_\parallel}{\gamma } \frac{\partial f}{\partial p_\parallel} , \label{eq:diff_eq_sol_gen}
\end{equation}
where $\eta=5-(p_\perp^2 + p_\parallel^2)/\gamma^2$ ($\eta$ ranges between $\eta=5$, in the non-relativistic limit $\gamma \simeq 1$, and $\eta=4$ in the relativistic limit $\gamma \gg 1$). From now on, we will assume the relativistic limit. We also note that $\mathbf{\nabla}_p \cdot\mathbf{F}_{rad} \propto p_\perp^2/\gamma$ indicates that compression of momentum space volume leads to bunching along the $p_\perp$ dimension. \textcolor{black}{In Eq. (\ref{eq:diff_eq_sol_gen}) two terms contribute to the evolution of the momentum distribution in the $p_\perp$ direction and are responsible for the cooling. The second term, $\left(p_\perp^3/\gamma\right) \partial f/\partial p_\perp$, is associated with the contraction of the momentum distribution domain, as can be seen from the momentum trajectory of a single particle $dp_\perp/dt = -2\alpha B_0 p_\perp^3/(3\gamma)<0$. On the other hand, the first term $\eta p_\perp^2 f/\gamma$ always provides a positive contribution to $\partial f/\partial t$ and assures particle number conservation.} 
 
We examine regions of momentum space where $p_\perp \sim \gamma$, \emph{i.e.} $ p_\perp \gg p_\parallel$ and $\gamma \gg 1$. Then a solution for Eq. (\ref{eq:diff_eq_sol_gen}) can be obtained by the method of characteristics:
\begin{equation}
    f(p_\perp, p_\parallel, t) = \frac{f_0\left(\frac{p_\perp}{1- \tau p_\perp}, \frac{p_\parallel}{1- \tau p_\perp }\right)}{\left( 1- \tau p_\perp \right)^4}. \label{eq:general_solution}
\end{equation}
where $\tau = \frac{2}{3} \alpha B_0 t$.
Eq. (\ref{eq:general_solution}) fully determines the temporal evolution for any given initial distribution $f_0$, and demonstrates several conclusions regarding the general evolution of momentum distributions undergoing synchrotron cooling. Firstly, The solution domain decreases with time, with an upper bound at $p_\perp^* = \tau^{-1}$ (such that $p_\perp \tau < 1$). Therefore, the distribution function is compressed within $p_\perp \leq p_\perp^*$, where $p_\perp^*$ describes the trajectory of a particle that $p_\perp(\tau = 0) = \infty$. 

\textcolor{black}{From Eq. (\ref{eq:general_solution}) we can conjecture that a Landau population inversion, characterized by $\partial f/\partial p_\perp >0$, develops in a finite time for a wide variety of initial momentum distributions (see supplementary material). This can be shown by rearranging $\partial f/\partial p_\perp >0$ and considering $p_\parallel = 0$  
\begin{equation}
    4\epsilon f_0\left( p_\perp',0\right)  >  -  p_\perp'\frac{\partial f_0}{\partial p_\perp'}\left( p_\perp',0\right), \label{eq:inequality_1}
\end{equation}
where $p_\perp' = p_\perp/(1-p_\perp \tau)$ and we have used the fact that $0<p_\perp<p_\perp^*$ to write $p_\perp = \epsilon p_\perp^* = \epsilon/\tau$, with $0\leq\epsilon<1$. For an initially stable distribution $f_0$ all terms in Eq. (\ref{eq:inequality_1}) are positive because $f_0>0$ and $\partial f_0 / \partial p_\perp \leq 0$, everywhere. Equation (\ref{eq:inequality_1}) illustrates a very simple condition for the development of unstable distributions. From this inequality, one can obtain the range of $p_\perp$ where the unstable region is developed. Moreover, furthering our conjecture, we have checked that a wide variety of momentum distributions fulfil Eq. (\ref{eq:inequality_1}), including Maxwellian, Maxwell-J\"uttner, constant negative slope distributions, power-laws up to the power of $4$, \textit{etc}. For all of these distributions, or combinations a population inversion \emph{i.e.} a ring momentum distribution, will be formed. }

In order to determine the relevant timescales for the ring formation and population inversion process, an isotropic Maxwellian distribution function is considered; more general distribution functions, such as a Maxwell-J\"uttner distribution function or Maxwellian beam distribution, will be studied numerically.
\begin{figure}[t]
    \centering
    \includegraphics[width=\linewidth]{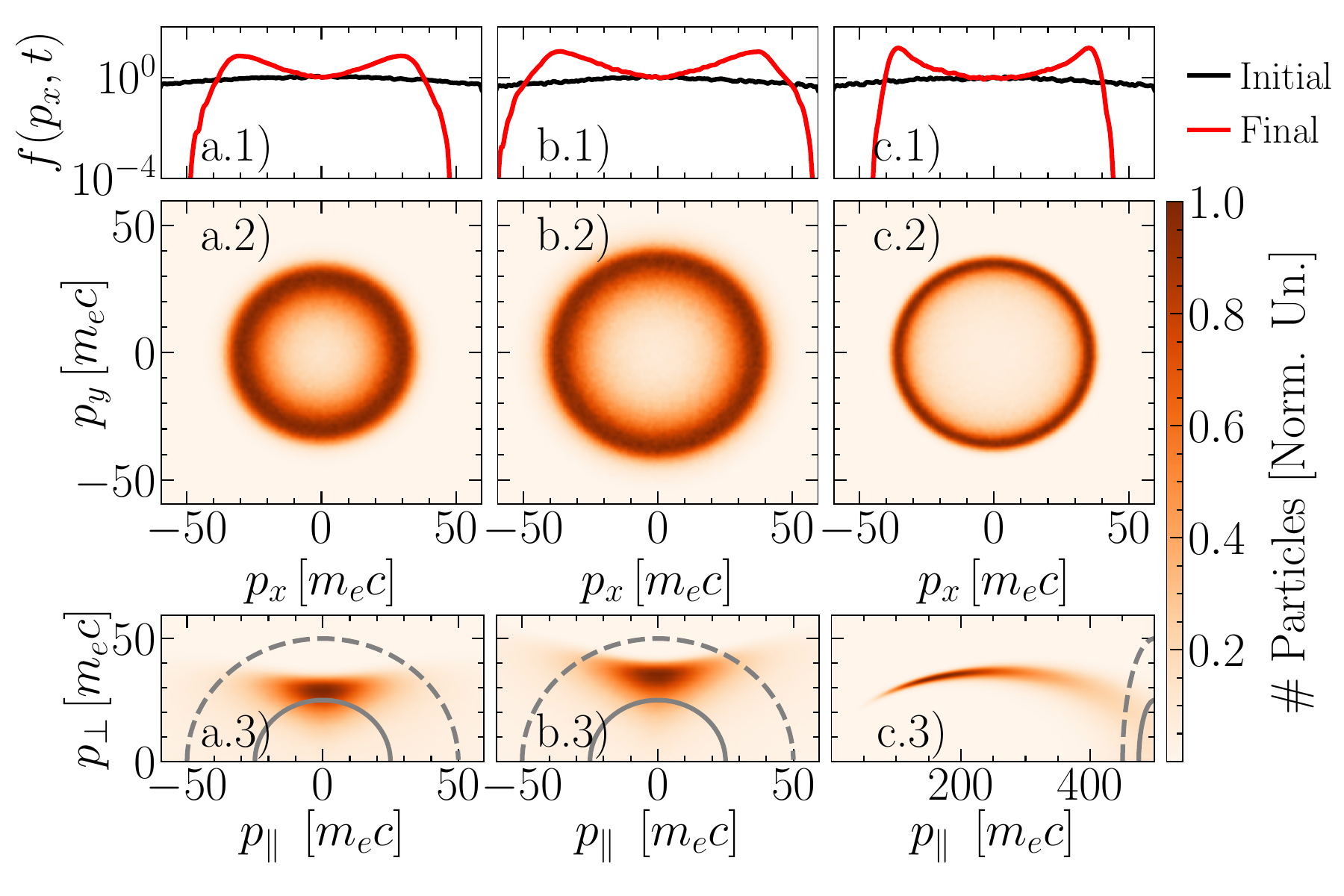}
    \caption{Particle-in-cell simulation results demonstrating the evolution of an initial isotropic Maxwellian distribution (column a), a Maxwell-J\"uttner distribution (b), and a beam with bulk $\gamma_b = 500$ and isotropic Maxwellian spread (c), at $t=3t_R$. For reference, the distribution function $f(p_x,p_y=0)$ is shown at $t=0$ and $t=3t_R$ on the top row (1). The second row (row 2) shows the perpendicular plane of the momentum distribution ($f_\perp(p_{x},p_{y})$, where $p_\perp^2 = p_{x}^2 + p_{y}^2$) and the bottom row (3) the $f(p_\perp, p_\parallel)$ momentum distribution, at $t=3t_R$. }
    \label{fig:rings}
\end{figure}
The initial Maxwellian distribution function, with thermal momentum spread $p_{th} = \sqrt{m_e k_B T}$, where $k_B$ is Boltzmann constant and $T$ is the temperature of the plasma, is defined as $f_{0 \, M}(p_\perp, p_\parallel)=e^{-(p_\perp^2 + p_\parallel^2)/(2p_{th}^2)}/(p_{th}^3(2\pi)^{1/2})$. The resulting ring radius $p_R(t)$ in momentum space, is defined as $\left|\partial_{p_\perp} f_\perp(p_\perp, t)\right|_{p_\perp =p_R(t)} = 0$, where we have defined $f_\perp$ as the integrated distribution along $p_\parallel$ \emph{viz.} $f_\perp(p_\perp, t) = \int_{-\infty}^{\infty} f(p_\perp, p_\parallel, t ) d p_\parallel$. Using Eq. (\ref{eq:general_solution}) to determine the temporal evolution of $f_{0 \, M}$, $p_R$ evolves as
\begin{equation}
    p_R (t) = \frac{1+ 6 p_{th}^2 \tau^2 - \sqrt{1+12 p_{th}^2 \tau^2}}{6 p_{th}^2\tau^3}, \label{eq:peakpos}
\end{equation}
For early times, $\tau=2 \alpha B_0 t/3  \ll 1$, the ring radius (in momentum space) grows linearly $p_R(t)\sim 2 p_{th}^2\alpha B_0 t$. At later times, $\tau \gg 1$, the radiation reaction cooling constricts the ring, reducing its radius as $p_R(t)\sim 3/(2\alpha B_0 t)$, increasing $f_\perp(p_R)$, and in turn, increasing $\partial_{p_\perp} f_\perp$ within $p_\perp < p_R$. The ring formation time $t_R$ is naturally defined as the transition point between these two regimes, \emph{i.e.} the time at which the ring stops growing and begins to decrease in momentum radius, defined as $\left|\partial_t  p_R(t)\right|_{t=t_R} = 0$, and given by 
\begin{equation}
    t_R = \frac{3}{ 4 \alpha B_0 p_{th}} = \frac{3}{ 4 \alpha \chi_\mathrm{th}} \label{eq:t_peak_max}
\end{equation}
where we have defined $\chi_\mathrm{th}$ (in normalized units) as $\chi_\mathrm{th}= B_0 p_{th}$. The ring formation time decreases as radiation reaction becomes more important (higher $\chi_\mathrm{th}$), and the timescale for the ring formation $t_R [\text{ns}]\simeq 7.5 B_0^{-2} \left[50\, \text{MG}\right]  p_{th}^{-1}\left[ 10 \, m_e c\right]$ is compatible with astrophysical and laboratory conditions.

These results can be generalized to particle beams since our calculations have considered the proper reference frame of the plasma/beam, where the fluid momentum of the beam $\overline{\mathbf{p}}=0$. For a beam propagating parallel to the magnetic field with Lorentz factor $\gamma_b$ all the previous results can be rescaled by the appropriate Lorentz transformations, $t_R = 3 \gamma_b / 4 \, \alpha \, \chi_\mathrm{th}$; in these conditions, an inverse Landau population is generated and evolves into a ring-beam distribution, \emph{i.e.} a beam with a pitch angle anisotropy. 

We have performed Particle-in-Cell (PIC) simulations with the PIC code OSIRIS \cite{fonseca2002osiris}, including classical \cite{vranic2016classical_m} and QED \cite{vranic2015particle, vranic2016classical_m} radiation reaction to confirm and to explore the theoretical findings. The full details of the simulation parameters are included in the Supplementary Material. Simulations with different initial distributions show the formation of the ring at $t=3t_R$, confirming the theoretical predictions for an initially isotropic Maxwellian distribution function $f_{0\, M}$ with $p_{th} = 50 \, m_e c$  (Fig. \ref{fig:rings}.a). Equivalent behavior is also evident for an initially isotropic Maxwell-J\"uttner distribution $f_0 \propto \exp(-\gamma/k_B T)$ with $k_B T = 50\ m_e c^2$ (Fig. \ref{fig:rings}.b). A beam with $\gamma_B = 500$ and a Maxwellian thermal spread $p_{th}=50 \, m_e c$ in the lab frame also evolved into a ring in the boosted timescale $t=3 \gamma_b t_R = 3\gamma_b / 4 \alpha \chi_\mathrm{th}$ (Fig.\ref{fig:rings}.c). \textcolor{black}{To demonstrate the ring formation in the weakly-relativistic regime, simulations with $p_{th} \sim 1\ m_e c$ were also performed (see supplementary material).}

\begin{figure}[t]
    \centering
    \includegraphics[width=\linewidth]{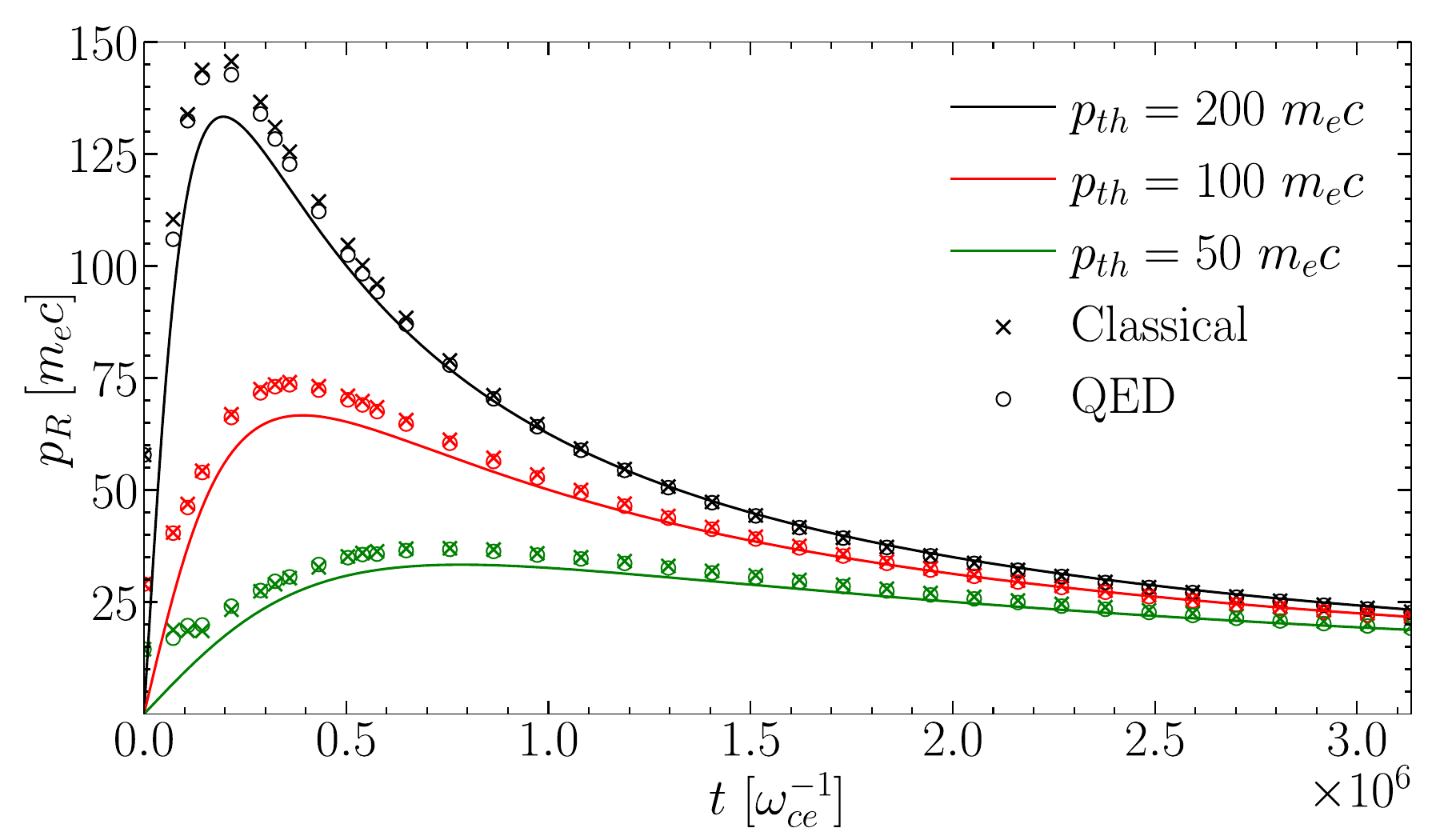}
    \caption{Temporal evolution of the ring radius $p_R$ for different initial conditions, showing the agreement between the theoretical predictions (line Eq. (\ref{eq:peakpos})) and simulations ($\circ$ classical and $\times$ QED) for different initial Gaussian distributions with momentum spreads $200,\ 100,\ \&\ 50\ m_e c$ (black, red, and green)}
    \label{fig:parameter_scan_spread}
\end{figure}
This evolution is further explored in Fig. \ref{fig:parameter_scan_spread}, focusing on initially Maxwellian distribution functions. The evolution of the  ring radius, Eq. (\ref{eq:peakpos}), shows excellent agreement with numerical simulations. Results with classical and QED radiation reaction are also shown. The average $\chi$ of the distribution function, $\bar{\chi}$, defined as $\bar{\chi}(t)=\int_0^\infty f_\perp(p_\perp, t) p_\perp^2 B_0 dp_\perp$ is a useful quantity to assess the importance QED radiation reaction. $\bar{\chi}$ decreases as the distribution function cools down (since $\mathbf{B_0}$ is constant). Thus, the maximum $\bar{\chi}$ for this configuration is always $\bar{\chi}(t=0) = \sqrt{\pi/2}\chi_\mathrm{th}$. In the simulations in Fig. (\ref{fig:parameter_scan_spread}), $\bar{\chi}(t=0) \sim 10^{-6} \ll 1$ and, as expected, in this regime the QED and classical results agree \cite{vranic2016quantum_m}. The discrepancy at early times in Fig. \ref{fig:parameter_scan_spread} between theory and simulations is due to the range of validity of the theoretical model ($p_\parallel \ll p_\perp$) -- outside this range the evolution of the distribution function $f$ deviates from the prediction of Eq. (\ref{eq:general_solution}); at later times, due to the differential cooling, most of $f$ is within the range of validity of the theoretical model, and a closer match between theory and simulations is observed.

We have explored the full quantum regime through simulations with $\chi_\textrm{th} = 0.25,\ 0.5\ \&\ 1$ by increasing the magnetic field strength (shown in the supplemental material). We explored this regime with both classical and QED radiation reaction simulations. Both simulations show that the rings are formed within similar timescales, the main difference is that the simulations \textcolor{black}{with quantum synchrotron emission} show rings with a larger width than the rings in the classical simulations (see supplemental material). Another fact to consider is that all distributions that are initially in the high $\bar{\chi} (t=0)$ regime eventually enter the classical regime $\bar{\chi} \ll 1$. This is expected, as $\bar{\chi} \simeq p_R B_0$, and $p_R$ decreases at late times as $p_R \propto t^{-1}$. The ring formation in the high $\chi_\textrm{th}$ regime and the transition to the classical regime will be studied in future work.

We have considered a simplified field configuration, demonstrating the role of differential cooling of radiation reaction to generate population inversion. This is a general property that should be observed for other field configurations. Other scenarios and field configurations where differential cooling can be relevant are associated with betatron oscillations in an ion channel \cite{wang2002x,lu2007generating, glinec2008direct, davoine2018ion} or direct laser acceleration configurations \cite{pukhov1999particle}; a population inversion is also expected in those conditions and this will be explored in future publications.

We have also performed simulations where the ring momentum distribution evolved for longer times, to assess the onset of the electron cyclotron maser instability (\textcolor{black}{ECMI}) by the inverted Landau population, as the ring momentum distributions are well-known to be kinetically unstable \cite{le1984direct,bingham2000generation, cairns2005cyclotron, cairns2008cyclotron, melrose2016cyclotron}. The growth rate for the fastest growing mode of the ECMI (the first harmonic of the X-mode), assuming a momentum distribution $f$ with small $p_\parallel$ spread, can be determined from the standard electron cyclotron maser theory \cite{le1984direct},  $ \Gamma(\omega) = - \frac{\pi}{4} \omega_{ce} \frac{\omega_p^2}{\omega^2} p_{re}^2 \left| \partial_{p_\perp} f(p_\perp, t)\right|_{p_\perp = p_{re}}$, where $p_{re}^2 =\frac{\omega_{ce}^2}{\omega^2} -1$ comes from the maser resonant condition and $\omega_p$ is the plasma frequency. Assuming $\partial f/\partial p_\perp  \sim \Delta f_\perp/\Delta p_\perp \sim f_\perp(p_R, t)/p_R$, $p_{re} \sim p_R$, with $p_R \gg 1$, then $\omega = \omega_{ce}/p_R$, and using the analytical results for an initially Gaussian $f_0$, the growth rate normalized with respect to $t_R$, at $t_R$, can be estimated as
\begin{equation}
    |\Gamma(t_R) t_R| \sim  0.5 \frac{\omega_p^2}{\omega_{ce}^2} p_{th}.
\end{equation}
For weakly magnetized scenarios, $\omega_p \gg \omega_{ce}$ (and a relativistically hot plasma such that $p_{th} \sim {\cal O} (1)$, the instability develops on time scales comparable (or shorter) to the ring formation time. In the opposite (and more interesting) limit, the ring will be formed and stable before the onset of the instability. An important fact to consider is that independently of $\omega_p / \omega_{ce}$ (and even for $\omega_p / \omega_{ce} \ll 1$), the ring will continue to constrict (and $\partial f/\partial p_\perp$ to increase), to the point where the growth rate is strong enough for the onset of the maser process and the emission of coherent radiation.

\textcolor{black}{In the limit of small magnetic fields $B\to0$ according to Eq. (\ref{eq:t_peak_max}) $t_R\to\infty$. In this scenario, effects that can inhibit the ring formation must be included in our analysis. That is the case of collisional processes (e-e, e-i collisions, pair annihilation, and Compton scattering \cite{lightman1982relativistic, goldston1995introduction}). The competition with these processes (when $t_R$ becomes comparable to their typical time scales) might inhibit the ring formation and the Landau population inversion. However,} for the range of conditions in the magnetospheres of compact objects, such as magnetars and pulsars, the ring formation time (from Eq.(\ref{eq:t_peak_max})) $t_R \simeq {\cal O}(\text{ps})$, considering GigaGauss field strengths and relativistic plasmas with $p_{th} \sim 100 \, m_e c$, is much shorter than all of the other timescales; for laboratory experiments with 10's of MG $B$ field strengths and relativistic $p_{th}\sim 10 \, m_e c$, the ring formation occurs in the nanosecond time-scale, which hints that other configurations might be more favorable to explore this process in the laboratory. 

It is generally accepted that coherent emission processes must be at play around compact objects \cite{melrose1995models, melrose2021pulsar}. Among these processes, the electron cyclotron maser instability requires a Landau population inversion \cite{le1984direct,bingham2000generation, cairns2005cyclotron, cairns2008cyclotron, melrose2016cyclotron}. Moreover, some of the other proposed coherent emission mechanisms assume strongly radiation cooled down beams where $p_\perp \ll m_e c$ \cite{lyutikov2021coherent}. We have demonstrated that in scenarios with strong radiation reaction cooling, transverse momentum distributions with inverted Landau populations are pervasive. These distributions are unstable and can drive coherent emission via kinetic plasma instabilities. Our analytical model shows excellent agreement with numerical simulations, demonstrating the relevance of this process in the classical and in the QED regimes of radiation reaction. We conjecture that our findings are also valid for other field configurations, namely in laboratory conditions, e.g. the focusing field of ion-channels in laboratory conditions with betatron oscillations \cite{wang2002x,glinec2008direct,davoine2018ion}, also opening the way to the laboratory exploration of Landau population inversion via radiation reaction cooling.

\begin{acknowledgments}
We would like to acknowledge enlightening conversations with Prof. R. Bingham,  Prof. A. R. Bell, Prof. M Lyutikov, Mr. R. Torres, Dr. T. Grismayer \& Dr. T. Silva. This work was supported by FCT (Portugal) (Grant UI/BD/151559/2021 and X-MASER - 2022.02230.PTDC) and the European Research Council (ERC) -2015-AdG Grant 695088 - InPairs. PIC simulations were performed at LUMI within EuroHPC-JU project EHPC-REG-2021R0038. \\
\textcolor{black}{After this letter was submitted for publication, we became aware of work by V.~Zhdankin, M.~W.~Kunz \& D.~A.~Uzdensky \cite{zhdankin2022synchrotron} also demonstrating that a collisionless, synchrotron-cooling plasma develops pressure anisotropy. Those authors focused on high-energy astrophysical systems in which the plasma beta is large enough for this pressure anisotropy to be unstable to the firehose instability.}\\
\end{acknowledgments}
\bibliography{apssamp}
\clearpage
\begin{widetext}
\textcolor{black}{\subsection{Relativistic momentum distribution undergoing synchrotron cooling develop $\partial f / \partial p_\perp > 0$ in a finite time.}
The radiation reaction force for a relativistic particle in a constant magnetic field $\mathbf{B}$ is
\begin{equation}
    \frac{d\mathbf{p}}{d \tau} = -\frac{p_\perp^2}{\gamma} \mathbf{p} ,
\end{equation} 
where $\tau= 2\alpha B_0/3$, as defined in the main text. We focus on particle trajectories with $p_\parallel = 0$ as the population inversion is expected to develop in that region of the distribution. We can calculate the trajectory of a particle in the perpendicular momentum space: 
\begin{equation}
    p_\perp (\tau, p_{\perp 0}) = \frac{p_{\perp 0}}{1+p_{\perp 0} \tau},
\end{equation}
where $p_{\perp 0}$ is the initial perpendicular momentum. For a particle at $p_{\perp 0} = \infty$ the trajectory simplifies to
\begin{equation}
    p_\perp^* (\tau) = \frac{1}{\tau}, \label{eq:full_traj}
\end{equation}
Eq. (\ref{eq:full_traj}) describes the trajectory of a particle cooling from $p_{\perp 0}= \infty$ at $\tau = 0$ to $p_\perp = 0$ at $\tau = \infty$. Therefore, for any distribution function cooling due to synchrotron radiation, its domain lies within $0<p_\perp< p_{\perp}^*$. Moreover, a momentum distribution undergoing synchrotron cooling also obeys $\left. \partial f/\partial \tau \right|_{p_\perp = 0} = 0$ because particles with no perpendicular momentum will not cool down, and the momentum distribution at $p_\perp = 0$ remains constant over time, as shown in the simulations (Fig. (1)).\\
Due to the conservation of the number of particles, the distribution function obeys
\begin{equation}
    N = \int_0^\infty f(p_\perp, \tau = 0)2 \pi p_\perp dp_\perp = \int_0^{p_{\perp}^*(\tau)} f(p_\perp, \tau) 2 \pi p_\perp dp_\perp, 
\end{equation} 
where $N$ is the total number of particles. We can rewrite the right-hand side as
\begin{equation}
    N =\int_0^{p_{\perp}^*(\tau)} f(p_\perp, \tau)2 \pi p_\perp dp_\perp= p_\perp^* f_{avg}(\tau),
\end{equation} 
where $f_{avg}$ is the average $f$ within $0<p_\perp< p_\perp^*$. Thus, using Eq. (\ref{eq:full_traj}),
\begin{equation}
    f_{avg} = \tau\int_0^{\infty} f(p_\perp, \tau)2 \pi p_\perp dp_\perp=\tau N.
\end{equation}
If the average of the distribution becomes larger than the value of the distribution at $p_\perp = 0$ this implies a region where $\partial f / \partial p_\perp > 0$. Thus,
\begin{equation}
    f(p_\perp = 0, \tau) <f_{avg} = \tau N.
\end{equation}
As $f(p_\perp = 0, \tau)$ is constant over time, and finite valued, there is a finite time at which this inequality becomes true. Therefore, a population inversion, i.e. a region where $\partial f/\partial p_\perp > 0$ occurs within a finite time.
}
\newpage

\subsection{Simulation parameters}
For the simulations, we have considered the same physical scenario examined analytically. There is in the $x_1$ direction a strong magnetic field $B_0 = 2.2\times10^{-6}$ (Normalized with respect to the Schwinger field $B_{sc}$) with an associated cyclotron frequency $\omega_{ce} = |e| B_{sc}B_0/m_e $, where $e$ is the electron charge and $m_e$ the electron mass. We normalised timescales and spatial dimensions with respect to the gyrofrequency $\omega_{ce}$ and $c/\omega_{ce}$, respectively. And momentum with respect to $m_e c$. The simulations employ a temporal resolution that guarantees the gyromotion is accurately resolved $\Delta t \sim 0.0099\ \omega_{ce}^{-1}$. The typical simulation is performed in one spatial dimension (and three momentum dimensions), using $5000$ cells and $1024$ particles per cell, with a spatial domain length along the $x_1$ direction of $L_x = 50\ c\omega_{ce}^{-1}$ with periodic boundary conditions. This yields a grid resolution of $\Delta x = 0.01\ c/\omega_{ce} = 0.99c \Delta t $, which verifies the Courant condition $\Delta x > c \Delta t$. A low-density electron plasma with plasma frequency $\omega_p = 10^{-4}\ \omega_{ce}$ fills the whole simulation domain with a background of immobile ions. Three different momentum distributions are initialized. A Maxwellian distribution $f_{0\, M} \propto e^{-(p_\parallel-p_\perp)^2 /(2 p_{th}^2)}$, with an isotropic momentum spread $p_{th}= 50\ m_e c$, a Maxwell-J\"uttner distribution $f_{0\, MJ} \propto e^{-\gamma m_e c/p_{th}}$ and a Maxwellian beam distribution $f_{0\, Mb} \propto e^{-((p_\parallel-\gamma_b m_e c)^2 -p_\perp^2)/(2 p_{th}^2)}$, where $\gamma_b$ is the bulk Lorentz factor of the beam, which was chosen to be $\gamma_b = 500$.

The macro-particles employ a cubic interpolation. We tested and compared different current smoothing filters, it was found that smoothing did not significantly affect the ring formation under these simulation conditions, for this reason, the final simulation setup employed a first-order binomial smoothing. The OSIRIS PIC code employs the reduced Landau-Lifshitz model (LLR) for classical radiation reaction, which includes the two leading orders of the full Landau-Lifshitz formulation, as described in \cite{vranic2016classical}.

For the parameters scan we kept all parameters constant and only changed the $p_{th}$ for the initial Maxwellian distribution. Employing $p_{th} = 50,\ 100\ \&\ 200 m_e c$.

\textcolor{black}{To confirm that the numerical heating/energy conservation is addressed properly we have compared energy conservation with and without radiation reaction for the same set of numerical parameters and calculated the radiated energy for all the runs presented in this study. This is shown in Fig. (\ref{fig:ene}), which accounts for the energy of the plasma particles, the energy radiated through synchrotron cooling and the expected kinetic energy of the plasma particles as predicted from our analytical results. In the bottom row of Fig. (\ref{fig:ene}) we show the evolution of the percentile change of the total energy, \textit{i.e.} the sum of the plasma and synchrotron radiated energy. Both plots demonstrate that energy is accurately conserved throughout the simulations, even for many time steps. 
\begin{figure}[h!]
    \centering
    \includegraphics[width=0.65\textwidth]{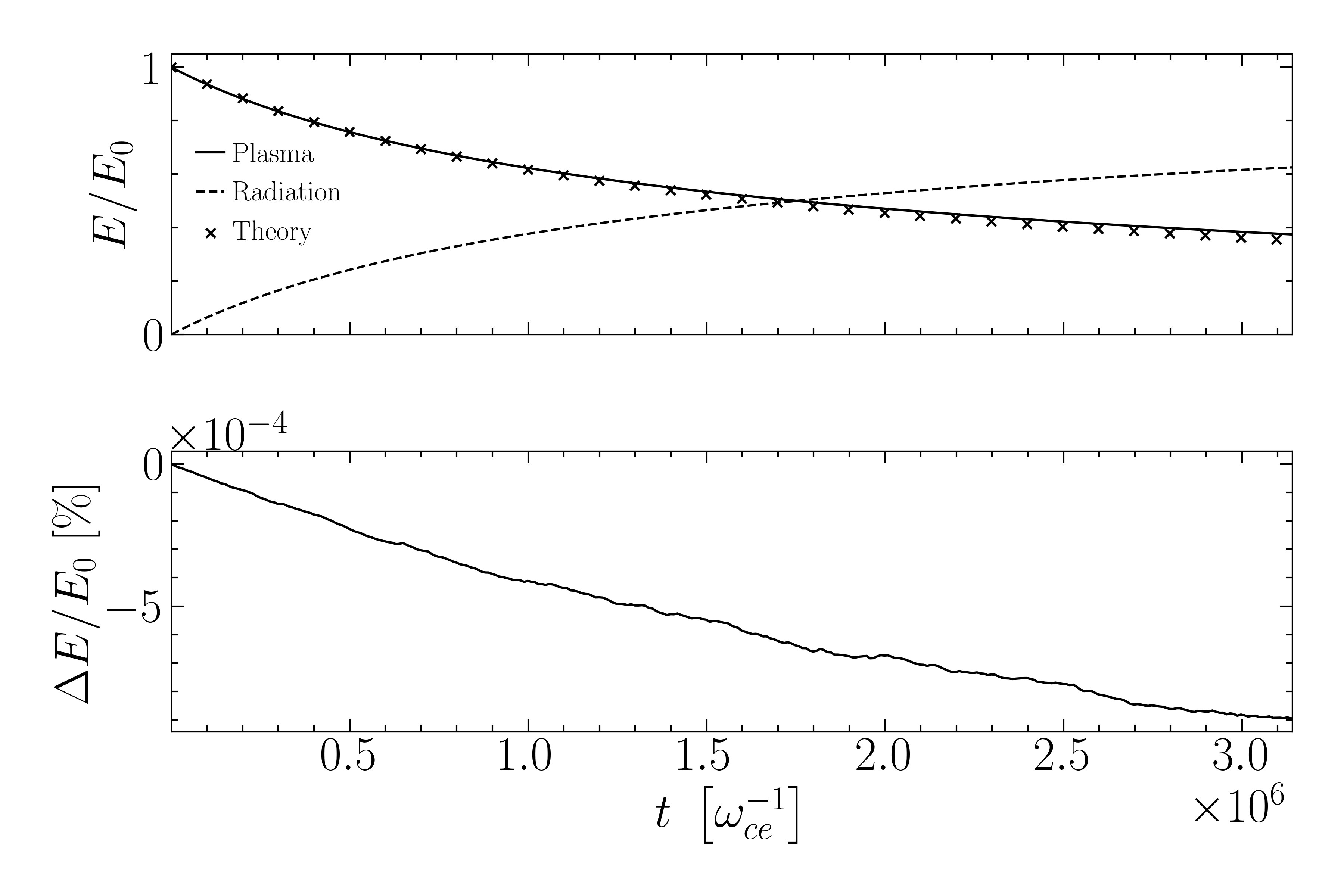}
    \caption{Top: Simulation results for the total energy for an initial isotropic Maxwellian plasma with $p_{th }= 50 \ m_e c$ and initial energy $E_0$ as a function of simulation time $t$. The plot shows plasma energy (continuous line), the energy from the synchrotron radiation diagnostic (dashed line) and the evolution of the energy of the plasma according to the theoretical calculation $E=\int d\mathbf{p}   f(p_\perp,p_\parallel, t) (\gamma-1) $ (crosses), where $f(p_\perp,p_\parallel,t)$ is given by Eq. (5). Bottom: Percentual change of the total simulation energy (sum of plasma energy and synchrotron radiation) over time.}
    \label{fig:ene}
\end{figure}}

\newpage
\textcolor{black}{\subsection{Simulation results for a weakly-relativistic distribution with $p_{th} = m_e c$}
Here we show the simulation results for an isotropic Maxwellian distribution with $p_{th} =  m_e c$, Most of the distribution lies in the region where $\gamma \simeq 1$, simulations with smaller $p_{th}$ significantly increase the timescale for the onset of the ring distribution. This result illustrates the development of the ring distribution beyond the range of validity of the theoretical model.
\begin{figure}[!h]
    \centering
    \includegraphics[width=0.5\linewidth]{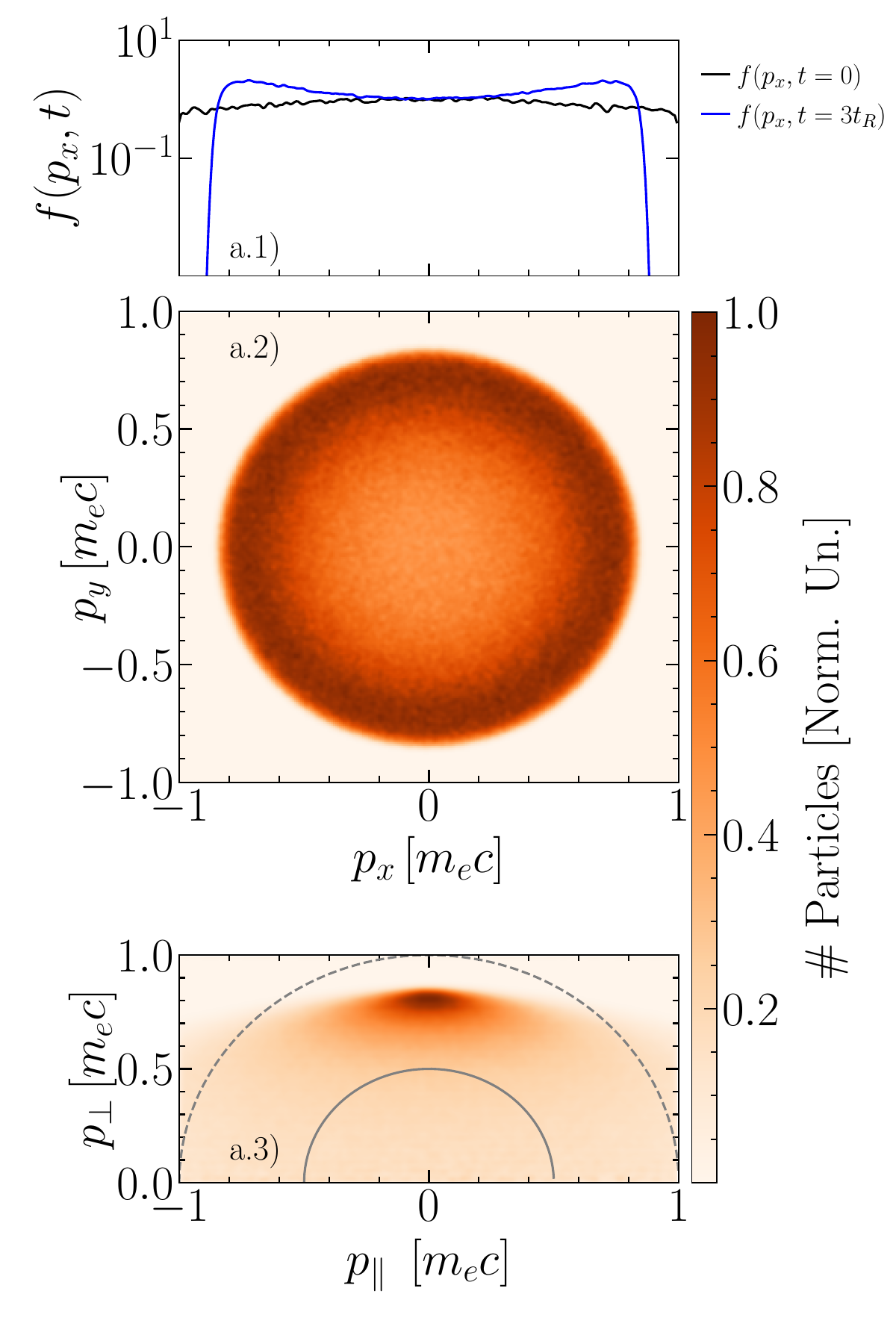}
    \caption{Particle-in-cell simulation results demonstrating the evolution of an initial isotropic Maxwellian distribution. For reference, the distribution function $f(p_x,p_y=0)$ is shown at $t=0$ and $t=3t_R$ on the top row (1). The second row (row 2) shows the perpendicular plane of the momentum distribution ($f_\perp(p_{x},p_{y})$, where $p_\perp^2 = p_{x}^2 + p_{y}^2$) and the bottom row (3) the $f(p_\perp, p_\parallel)$ momentum distribution, at $t\sim t_R$.}
    \label{fig:sup_small}
\end{figure}
}

\newpage
\subsection{Ring distributions in the QED regime $\bar{\chi}(t=0)\sim 1$}
Simulations with $B_0 = 0.005$ (Normalised with respect to the Schwinger field $B_{Sc}$), for different values of $p_{th}$, such that $\bar{\chi}(t=0)=0.25,\ 0.5\ \&\ 1$ are presented here. The simulations employ the QED and classical radiation reaction solvers and develop into ring momentum distributions within similar timescales. The key difference is that the ring momentum distributions including QED synchrotron emission have a larger ring width.
\begin{figure}[!h]
    \centering
    \includegraphics[width=0.78\linewidth]{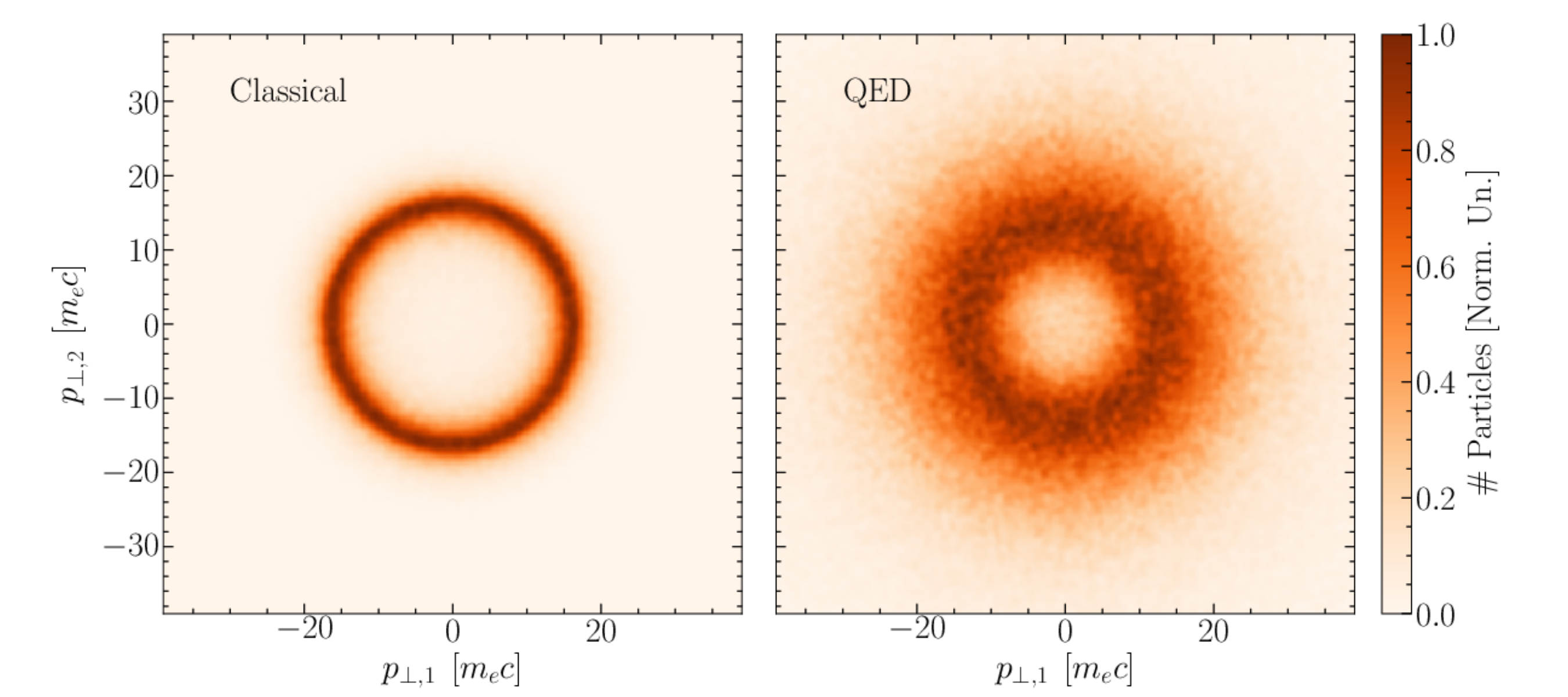}
    \caption{Evolution of an initial isotropic Maxwellian distribution with $p_{th} = 50 m_e c$ at $t=3 \, t_R$. The simulations employ a magnetic field strength $B_0 = 0.005$ (normalized to the Schwinger field). Thus, $\chi_{th} = 0.25$.
    The momentum distribution is shown integrated over the magnetic field direction, demonstrating the formation of the ring for both the classical and QED solver in the same timescale.}
    \label{fig:parameter_scan_spread_high_chi}
\end{figure}
\begin{figure}[!h]
    \centering
    \includegraphics[width=0.78\linewidth]{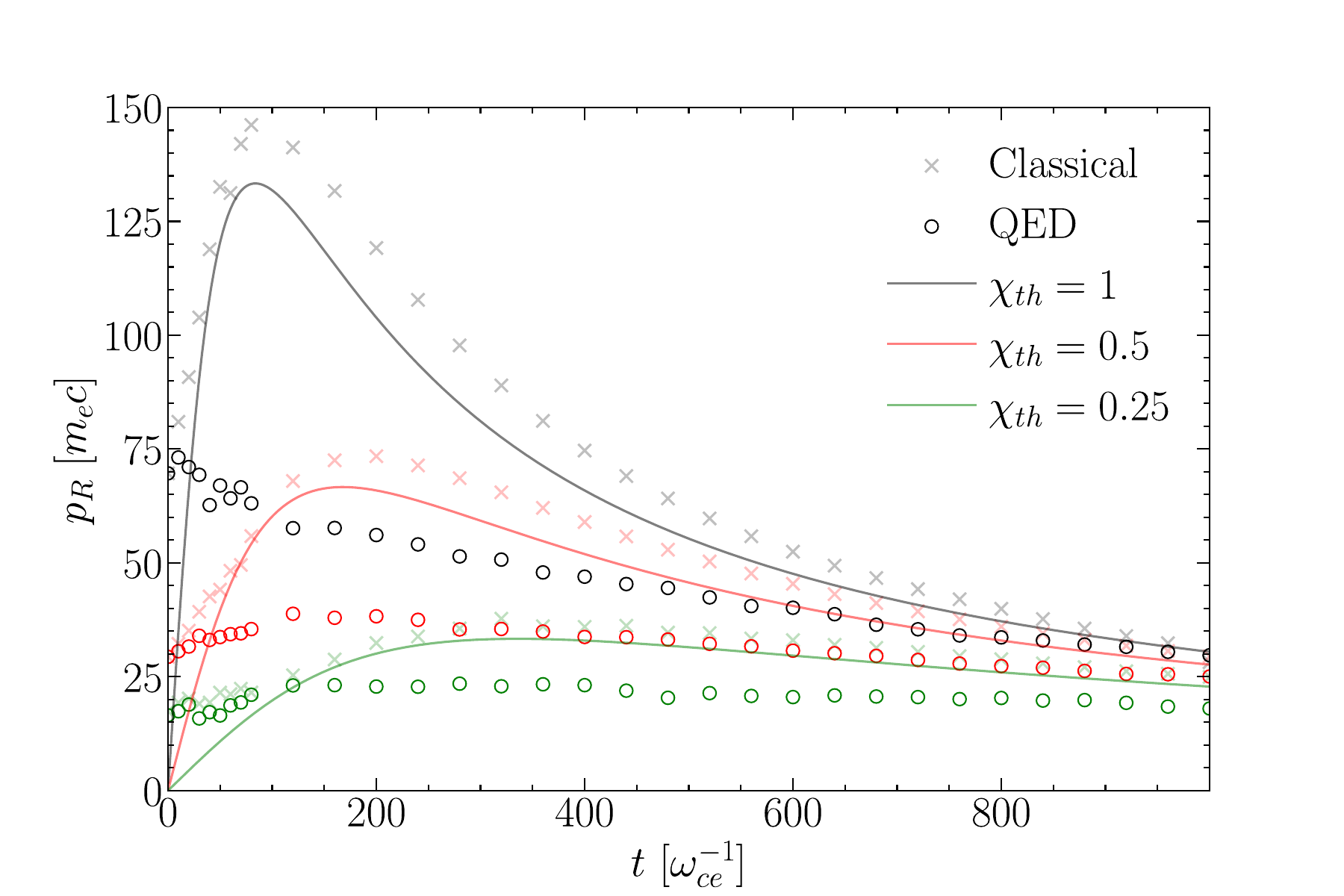}
    \caption{Same as Fig. 2 in the paper with $B_0 = 10^{-2}$ so $\bar{\chi}(t=0)=0.5,\ 1\ \&\ 2$. The lines describe  the analytical results for the ring radius evolution as predicted for the low $\chi$ regime. The QED-dominated regime still produces the ring momentum distribution within a comparable timescale as the classical radiation reaction. The ring evolution is altered, the rings are wider and they do not grow in radius as much as in the classical case.}
    \label{fig:parameter_scan_spread_high_chi_sup}
\end{figure}
\providecommand{\noopsort}[1]{}\providecommand{\singleletter}[1]{#1}%

\end{widetext}
\end{document}